\documentclass{PoS}

\usepackage[nolist,nohyperlinks]{acronym}

\usepackage{cleveref}
\usepackage{subcaption}
\usepackage{wrapfig}
\usepackage{siunitx}
\usepackage{multicol}
\usepackage{overpic}

\newcommand*\elem[2]{${}^{#1}\mathrm{#2}$}

\usepackage{lipsum} % sample text

\title{GAPS: Searching for Dark Matter using Antinuclei in Cosmic Rays}

\ShortTitle{GAPS Overview}

\author{\speaker{R.~Bird}\\
University of California, Los Angeles, Los Angeles, CA 90095\\
E-mail: \email{ralphbird@astro.ucla.edu}}
\author{for the GAPS collaboration\footnote{for collaboration list see PoS(ICRC2019)1177}}

\abstract{The General Antiparticle Spectrometer (GAPS) will carry out a sensitive dark matter search by measuring low-energy ($\mathrm{E} < \SI{0.25}{GeV/nucleon}$) cosmic ray antinuclei. The primary targets are low-energy antideuterons produced in the annihilation or decay of dark matter. At these energies antideuterons from secondary/tertiary interactions are expected to have very low fluxes, significantly below those predicted by well-motivated, beyond the standard model theories. GAPS will also conduct low-energy antiproton and antihelium searches. Combined, these observations will provide a powerful search for dark matter and provide the best observations to date on primordial black hole evaporation on Galactic length scales.

The GAPS instrument detects antinuclei using the novel exotic atom technique. It consists of a central tracker with a surrounding time-of-flight (TOF) system. The tracker is a one cubic meter volume containing 10 cm-diameter lithium-drifted silicon (Si(Li)) detectors. The TOF is a plastic scintillator system that will both trigger the Si(Li) tracker and enable better reconstruction of particle tracks. After coming to rest in the tracker, antinuclei will form an excited exotic atom. This will then de-excite via characteristic X-ray transitions before producing a pion/proton star when the antiparticle annihilates with the nucleus. This unique event topology will give GAPS the nearly background-free detection capability required for a rare-event search.

Here we present the scientific motivation for the GAPS experiment, its design and its current status as it prepares for flight in the austral summer of 2021-22.
}

\FullConference{36th International Cosmic Ray Conference -ICRC2019-\\
		July 24th - August 1st, 2019\\
		Madison, WI, U.S.A.}

\begin{document}

\section{Introduction}
\label{sec:Intro}
The \ac{GAPS} is a balloon-borne experiment to study low-energy ($<\SI{0.25}{GeV/nucleon}$) antinuclei in cosmic rays with the first of three proposed long-duration balloon flights from McMurdo station in the austral summer of 2021-2. Using the novel exotic atom technique, it will provide a new, complementary method to the magnetic spectrometer measurements conducted by AMS-02 \cite{ams} and BESS \cite{bess}, whilst increasing the sensitivity and lowering the energy range of these earlier searches.

The primary goal of \ac{GAPS} is to search for antideuterons.  At these low energies the predicted flux of antideuterons from secondary and tertiary sources (primary cosmic ray protons interacting with the \ac{ISM}, \cref{sec:Antinuclei}) is expected to be very low, well below what is detectable using current technology.  However, as was first discussed more than 15 years ago \cite{donato}, well-motivated dark matter models predict a low-energy antideuteron flux that is several orders of magnitude above this background, and within reach of the \ac{GAPS} experiment.  Combined with the tantalizing hints of excesses in antiprotons \cite{cuoco2016,cui2016} and in gamma-rays from the Galactic center \cite{ackermann2017}, antideuterons could provide the low-background, clean signal that would identify dark matter for the first time.

Recent years have also seen candidate cosmic ray antihelium events presented by the AMS-02 collaboration \cite{amsAHe}. \ac{GAPS}, operating using a different technique and with a lower energy threshold will provide a complimentary search.  As well as verifying any signal, the increased energy range will help in the understanding of the origin of these unexpected events.

\section{Low-Energy Antinuclei as Signals of New Physics}
\label{sec:Antinuclei}
\begin{figure}
\centering
\begin{minipage}{.48\textwidth}
    \centering
    \includegraphics[width=\textwidth, trim={0 0 0 0.5cm},clip]{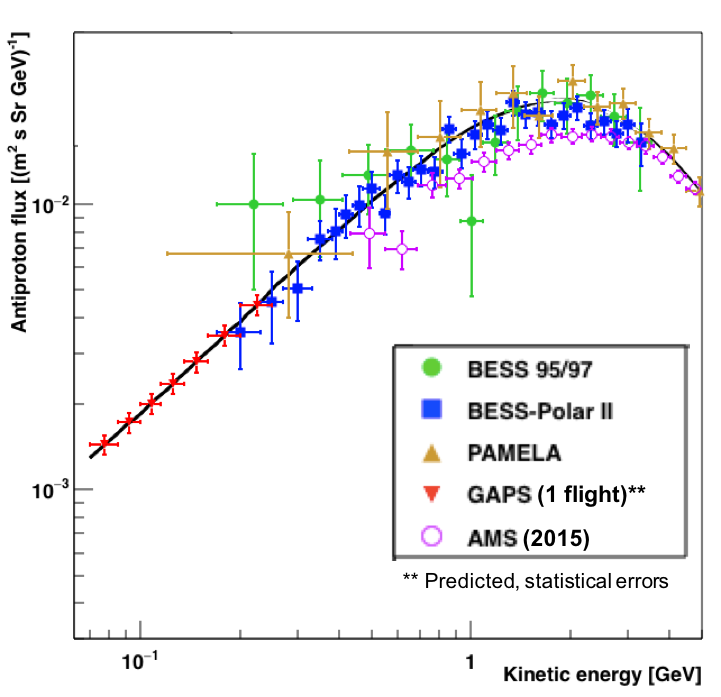}
    \caption{The expected antiproton spectrum measured by \ac{GAPS} in a single (30 day) flight (fig. adapted from \cite{aramaki2014}).}
    \label{fig:antiprotonSpec}
    \end{minipage}
\hfill
\begin{minipage}{.48\textwidth}
    \centering
    \includegraphics[width=\textwidth]{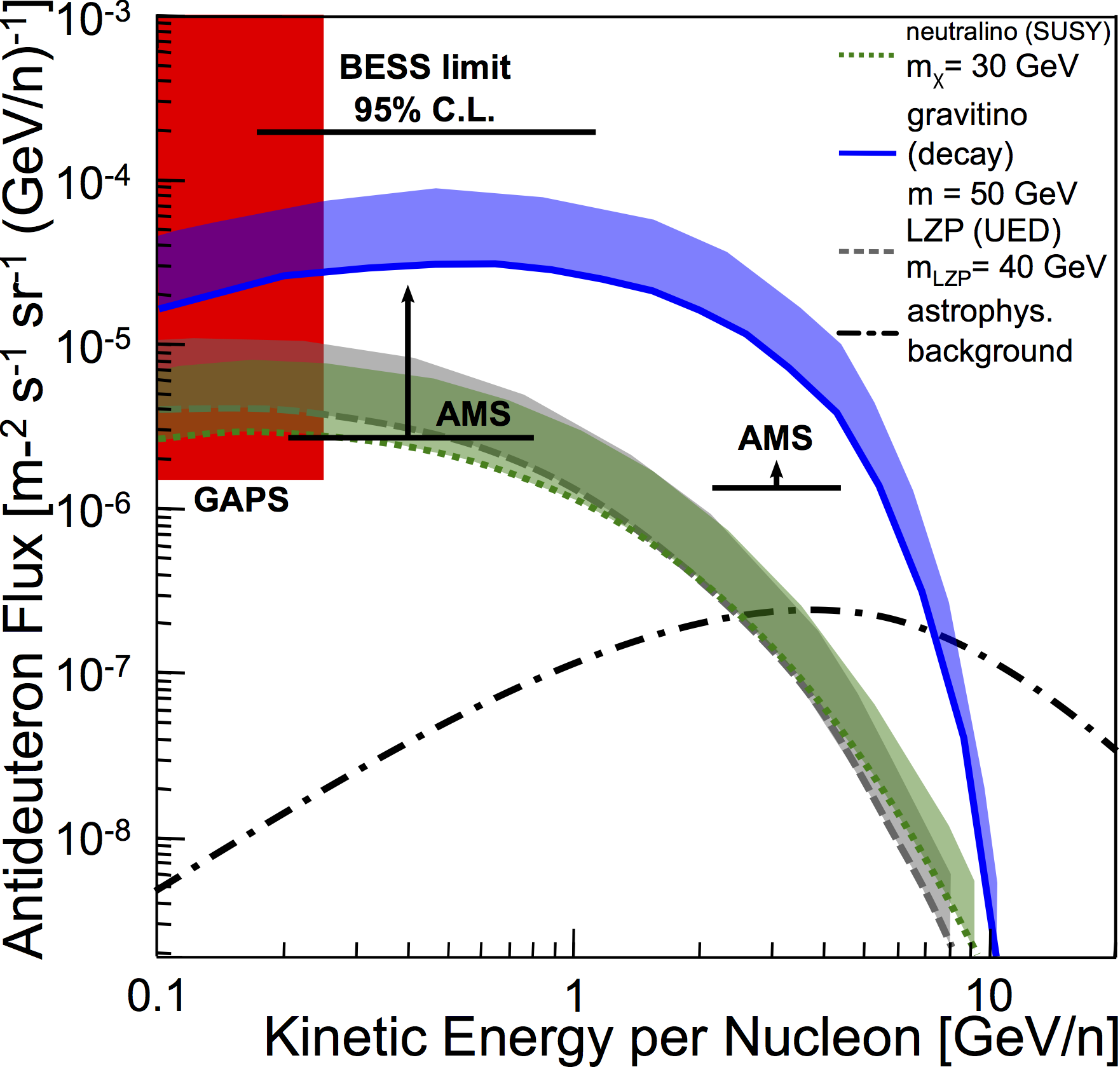}
    \caption{The predicted low energy antideuteron sensitivity of \ac{GAPS} (3$\sigma$ discovery limit) of three 35-day GAPS flights and five years of AMS-02 operation (fig. from \cite{dec2020}).}
    \label{fig:antideuteronSpec}
\end{minipage}
\end{figure}

As was discussed in \cite{donato,korsmeier2017,dec2020}, at high energies ($\gtrsim \SI{10}{GeV/nucleon}$) secondary antinuclei can be produced through the collision of high energy cosmic rays with the stationary \ac{ISM}.  However, at lower energies ($\lesssim \SI{1}{GeV/nucleon}$) this is heavily suppressed due to the kinematic requirements.  Likewise, tertiary signals from the down-scattering of high-energy antinuclei are heavily suppressed due to the instability of the antinuclei.  However, beyond the standard model physics may be able to produce low-energy antinuclei in appreciable numbers, a nearly background-free signal.  Dark matter annihilation provides a particularly interesting channel for the potential production of low-energy antinuclei.  Without the kinematic requirements of secondary production it is much easier to produce low-energy particles, and thus a variety of dark matter models predict an antideuteron flux that is several orders of magnitude higher than the secondary background (\cref{fig:antideuteronSpec}).  

In addition to searching for heavier antinuclei (antideuterons and antihelium), \ac{GAPS} will make a high-precision measurement of low-energy antiprotons (\cref{fig:antiprotonSpec}) \cite{aramaki2014}.  This will both be useful to search for deviations from the expected flux from secondary production and also to understand the varying effects of solar modulation with charge and mass at these energies (see \cite{muni2019} for a discussion of the latest results from the PAMELA experiment).

\section{The Exotic Atom Technique and the GAPS Concept}
\begin{figure}
\begin{minipage}{.48\textwidth}
    \centering
    \includegraphics[width=\textwidth]{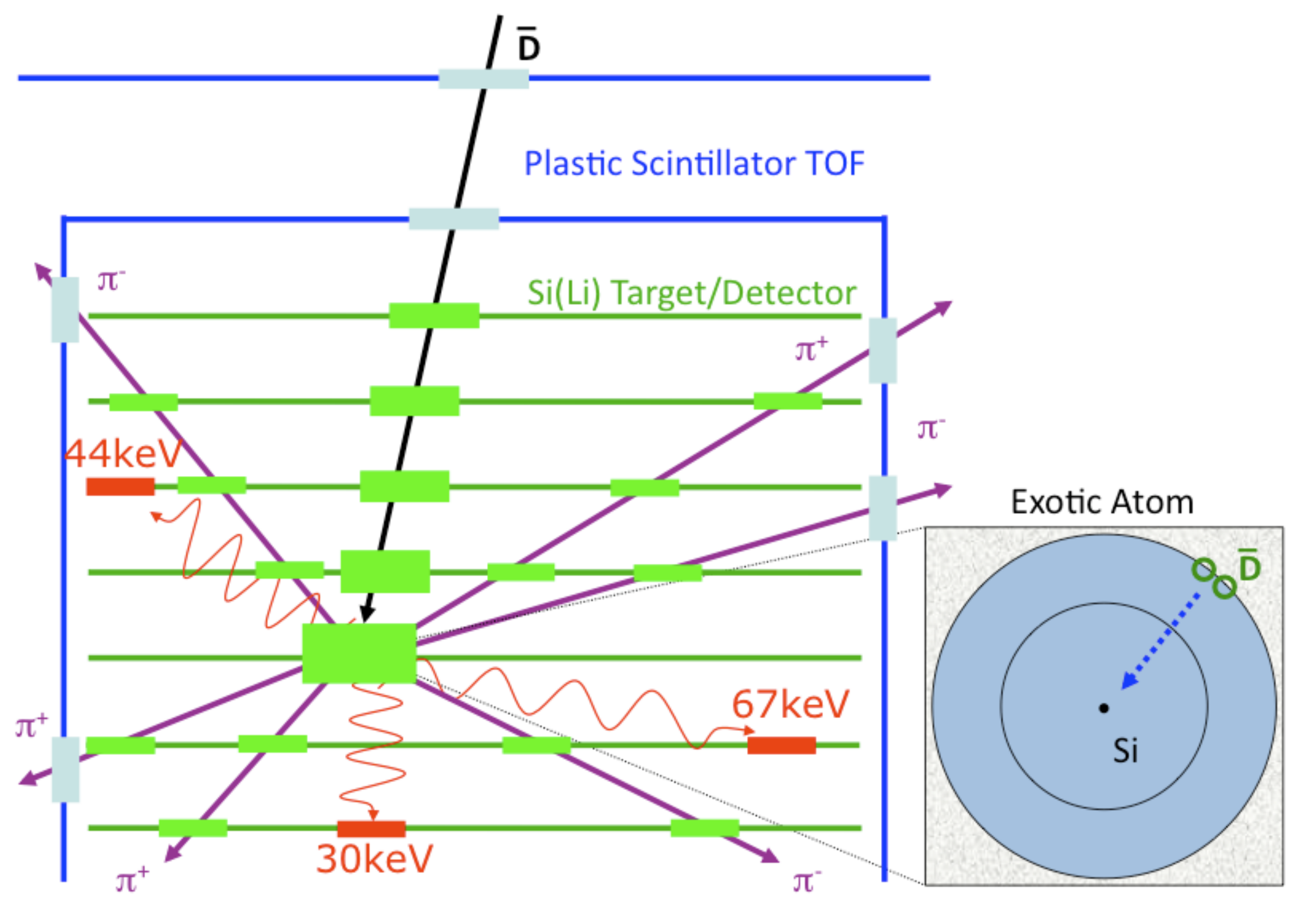}
    \caption{After slowing down and stopping in the detector, the antideuteron ($\overline{\mathrm{D}}$) forms an exotic atom with the silicon in the detector, before de-exciting with the release of characteristic X-rays and annihilating to produce pions and protons (fig. from \cite{aramaki2016}).}
    \label{fig:exoticAtom}
\end{minipage}%
\hfill
\begin{minipage}{.48\textwidth}
    \centering
    \includegraphics[width=\textwidth]{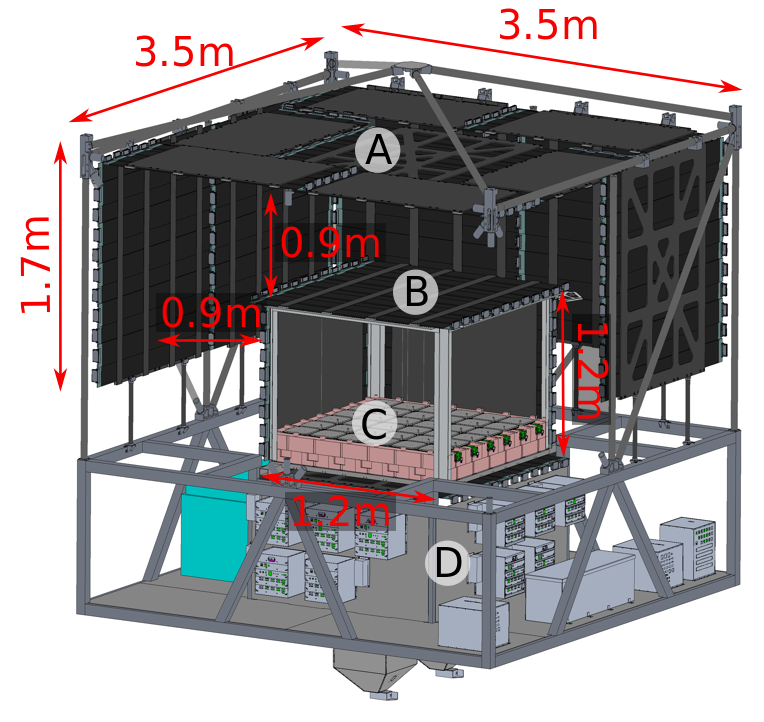}
    \caption{The \ac{GAPS} detector, the central tracker (C) is surrounded by the inner (``cube'', B) and outer (``umbrella'', A) \ac{tof} layers. The readout electronics, flight computer, ballast and other support infrastructure are located underneath the tracker (D).  Solar panels, cooling systems, antennae and thermal insulation are not shown for clarity.}
    \label{fig:GapsDetector}
\end{minipage}%
\end{figure}

\ac{GAPS} is designed to detect and characterize antinuclei using the exotic atom technique (\cref{fig:exoticAtom}) \cite{aramaki2016}, a new and powerful method that is complementary to the magnetic spectrometer methods employed by previous experiments.  Stopping antinuclei will form exotic atoms that decay on nanosecond timescales.  During the de-excitation process atomic X-rays will be produced, followed by a ``star'' of pions and protons from the nuclear annihilation.  Measuring the X-ray energies, the pion and proton multiplicities, and the stopping depth, velocity and energy deposition of the primary will allow \ac{GAPS} to fully characterize the primary particles.  This provides the required rejection of protons and other nuclei to search for an antinuclei search, and discrimination between the different antinuclei species and their isotopic separation.

Designed for low energy ($<\SI{0.25}{GeV/nucleon}$) antinuclei, \ac{GAPS} (\cref{fig:GapsDetector}) employs a central Si(Li) tracker (\cref{sec:track}) that stops the incident (primary) particles, is the site of the exotic atom products and also detects the produced pions, protons and X-rays.  This is surrounded by a plastic scintillator \ac{tof} (\cref{sec:tof}) system consisting of two layers separated by at least \SI{0.9}{m}.  The central layer fully encloses the tracker and the outer layer covers the upper half of the detector.   Together, these two systems will measure the velocity and charge of the primary particles, and detect the de-excitation X-rays and annihilation.  

\section{The GAPS Instrument : Design and Current Status}

\subsection{Tracker}
\label{sec:track}
The central tracker for \ac{GAPS} will be constructed of lithium-drifted silicon (Si(Li)) detectors to produce a large-area, high-temperature (\SIrange{-35}{-45}{\celsius}) detector system sensitive to both X-rays and ionizing particles.  The tracker is constructed of detector modules containing a $2\times2$ arrangement of \SI{10}{cm} diameter, \SI{2.5}{mm} thick, 8 strip detectors readout by a custom \ac{asic}. There will be ten layers of 36 detector modules spaced \SI{10}{cm} apart containing 1440 detectors and providing $\sim \SI{9}{m^2}$ of effective detection area.  Combined with the custom \ac{asic}, the requirements are an energy resolution of \SI{4}{keV} \ac{fwhm} in the energy range \SIrange{20}{100}{keV} and a dynamic range of \SIrange{0.01}{100}{MeV}.

\subsubsection{Si(Li) Detectors}
\label{sec:sili}
\begin{figure}
\begin{subfigure}[t]{0.32\textwidth}
\includegraphics[width=\textwidth,trim={0.5cm 0 10cm 0cm},clip]{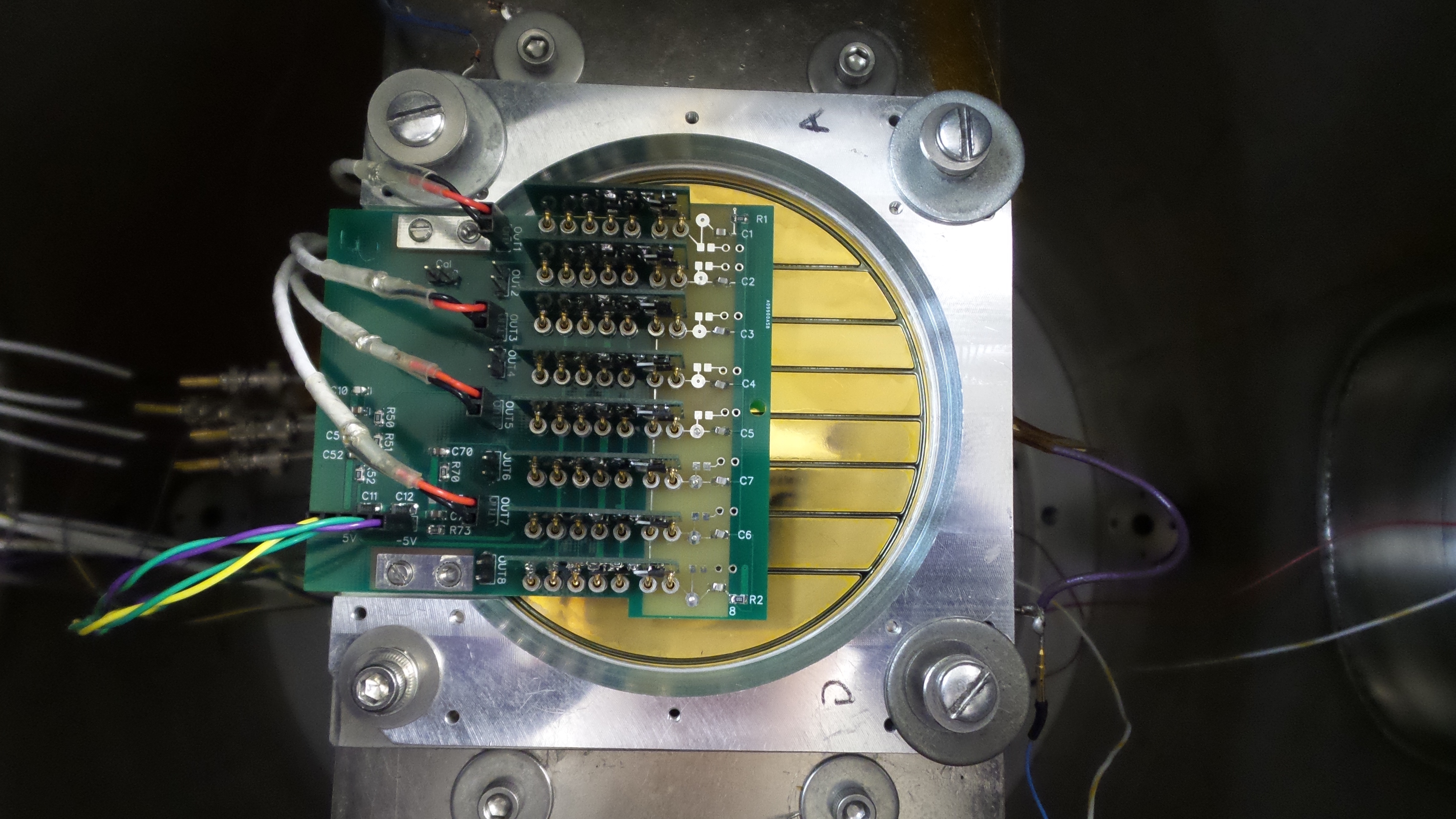}
\caption{An 8-strip Si(Li) detector mounted for testing in a  discrete-component preamplifier readout (as used to obtain the data presented in \cref{fig:SiLiResolution}).}
\label{fig:SiLiTesting}
\end{subfigure}
\hfill
\begin{subfigure}[t]{0.32\textwidth}
\includegraphics[width=\textwidth,trim={2.5cm 0 3cm 3cm}, clip]{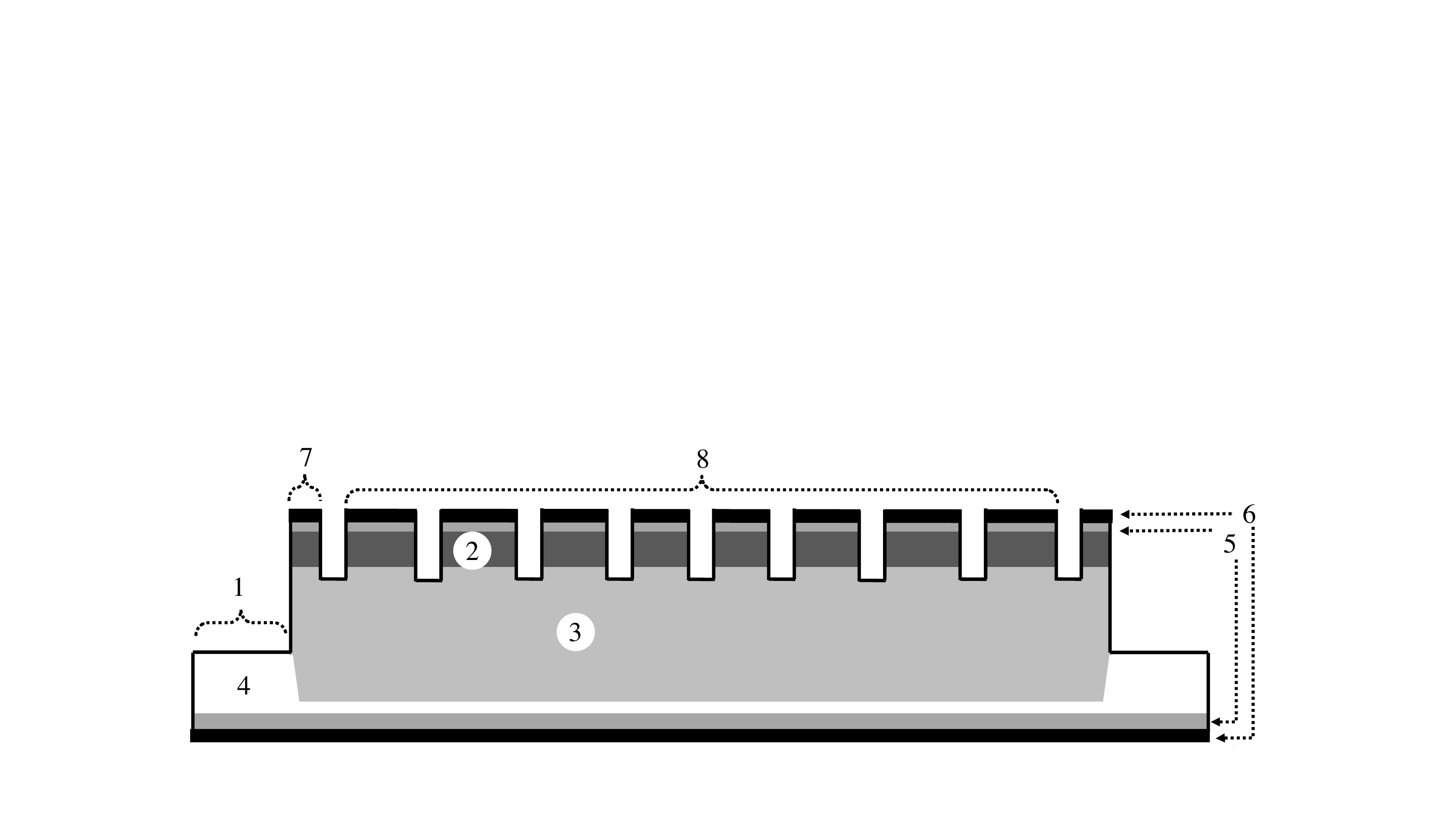}
\caption{A schematic of a Si(Li) detector, showing the top-hat structure, 8 equal-area strips, surrounding guard ring and sensitive volume. Details of labels and dimensions can be found in \cite{rogers2019}.}
\label{fig:SiLiSchematic}
\end{subfigure}
\hfill
\begin{subfigure}[t]{0.32\textwidth}
\includegraphics[width=\textwidth]{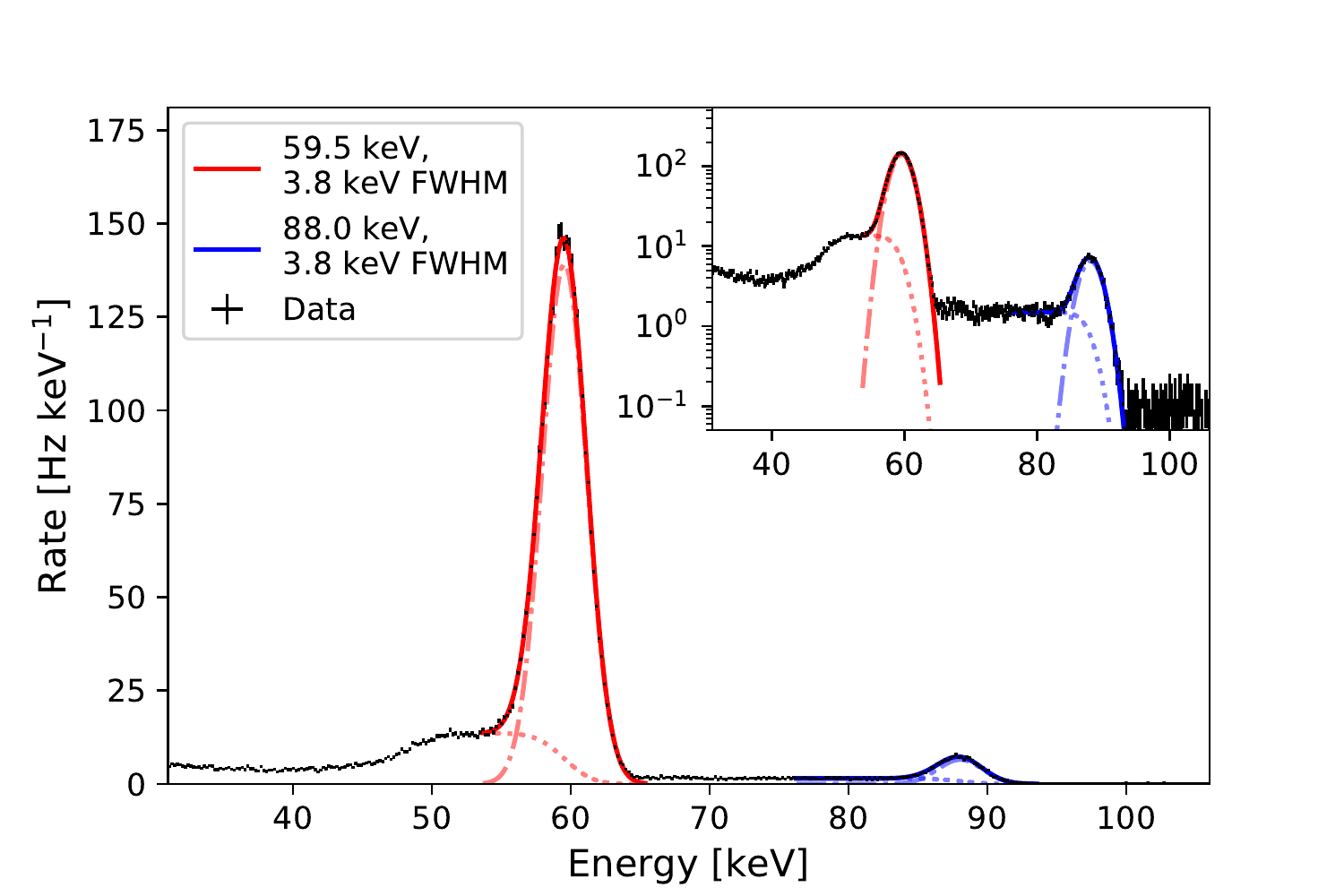}
\caption{A calibration spectrum for a prototype 4-strip detector showing the \SI{59.5}{keV} peak of \elem{241}{Am} and the \SI{88.0}{keV} peak of \elem{108}{Cd} with both peaks fitted independently. Lower-energy X-rays from \elem{108}{Cd} are mostly absorbed in the material between the source and the detector. \cite{rogers2019}}
\label{fig:SiLiResolution}
\end{subfigure}
\label{fig:SiLi}
\end{figure}

GAPS presents unique challenges for Si(Li) detector construction, operation and performance.  
The large detector area, low noise, and high temperature requirements have been met by developing a new fabrication technique, as described in detail in \cite{rogers2019} and \cite{kozai2019}. These detectors have been designed with a top-hat structure containing eight equal-area strips surrounded by a guard ring (\cref{fig:SiLiSchematic}).  Following fabrication, the detector is cleaned and passivated using VTEC PI-1388 Polyimide to protect the detectors from moisture and organic contaminants. 
Measurements using discrete-component preamplifier readout (\cref{fig:SiLiTesting}) have demonstrated energy resolution in the range of \SIrange{3}{4}{keV} \ac{fwhm} at \SI{59.5}{keV} and \SI{-35}{\celsius}, consistent with that expected based on the noise characteristics of the individual detectors \cite{rogers2019} (\cref{fig:SiLiResolution}). The passivation method has been separately validated (paper forthcoming).

\subsubsection{Tracker Electronics}
\label{sec:tracker}
To read out the 1,440 Si(Li) detectors with the required energy resolution and the dynamic range required the development of a custom \ac{asic}, known as the \ac{slider}.  In addition, the tracker electronics must power the Si(Li) detectors and provide interfaces to the flight computer.  Full details of the custom electronics are presented in \cite{slider2019}.

To achieve the desired range and resolution, the analog signal is processed using a low-noise charge-sensitive amplifier featuring dynamic signal compression.   This is achieved by taking advantage of a the non-linear features of the MOS capacitor in the feedback loop of the charge-sensitive preamplifier itself.  The signal is then shaped to one of eight peaking times ranging between \SI{250}{ns} and \SI{2}{\micro s}.  The signal is then split for readout via a sample-and-hold and for self-trigger operation.  

The power electronics sub-system provides high-voltage power to the Si(Li) detector modules and low-voltage power to the front-end \ac{asic} boards. The typical bias voltage of the Si(Li) detectors varies in the range \SIrange{150}{300}{V}, with a \SI{1}{V} accuracy. 

A prototype \ac{slider} \ac{asic} with a reduced number of readout channels (SLIDER8) has already been constructed and tested, a full flight prototype (pSLIDER32) is currently in production.

\subsection{Time-Of-Flight}
\label{sec:tof}
The GAPS \ac{tof} consists of two layers of plastic scintillator (Eljen EJ-200) which surround the central tracker to measure primary $\beta$ ($=v/c$; $v=$ velocity, $c=$ speed of light), detect annihilation products exiting the tracker volume to improve event reconstruction, and estimate the primary charge.  The \ac{tof} is constructed using counters of thickness \SI{6.35}{mm}, width \SI{16}{cm} and lengths ranging from \SIrange{1.2}{1.8}{m} with the light output detected at both end using six, $\SI{6}{mm}\times\SI{6}{mm}$ Hammamatsu S13360-6050VE \acp{sipm} mounted on a custom preamplifier board.  The inner ``cube'' consists of 64 counters providing near 100\% hermeticity and covering an area of \SI{15}{m^2}. Surrounding this, with a gap of \SI{0.9}{m} and covering all tracks with an incident zenith angle of less than \ang{60} is an ``umbrella'' of 132 counters covering an area of \SI{38}{m^2}.

The signals from the \acp{sipm} pass through an analogue front end which outputs two different signal channels, a high gain signal for event timing and light output measurement and a low gain signal for the trigger.  The trigger signal passes through a multi-level level trigger system.  A local trigger has five discriminator levels to estimate the energy deposited in each counter by the particle and transmits that information to a master trigger.  This uses the number of counters triggered at each level and the timing between the triggering signals to select potential low $\beta$ antinuclei events.  For events which receive a trigger the full trace will be sampled and digitized using a DRS4 \cite{drs4} based readout board and the event will then be processed by a central computer prior to storage or telemetry to ground.

The primary goal for the \ac{GAPS} \ac{tof} is to provide an accurate measurement of the primary particles' $\beta$, to meet the requirements needs a timing resolution ($\delta t = |t_A - t_B|/\sqrt{2}$) of \SI{500}{ps}.  Using 11,577 vertical muon events passing through the center of a \SI{1.8}{m} counter, $\delta t$ has been measured at $\SI{340}{ps} \pm \SI{2}{ps}$, significantly surpassing the design requirements. Further detail on the \ac{GAPS} \ac{tof} can be found in \cite{quinn2019}.

\begin{figure}
\begin{subfigure}[t]{0.66\textwidth}
\includegraphics[width=\textwidth,trim={2.5cm 0 3cm 3cm}, clip]{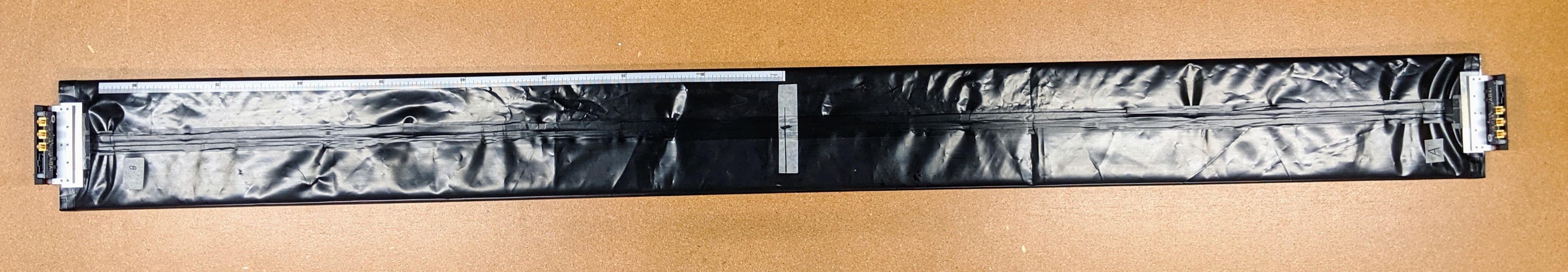}
\caption{Prototype \SI{1.8}{m} long counter used in lab testing. The preamplifiers have been mounted, and the main body sealed. This preamplifier mounting variant is not light tight, so a dark box is used for testing.}
\label{fig:Paddle}
\end{subfigure}
\hfill
\begin{subfigure}[t]{0.32\textwidth}
\includegraphics[width=\textwidth]{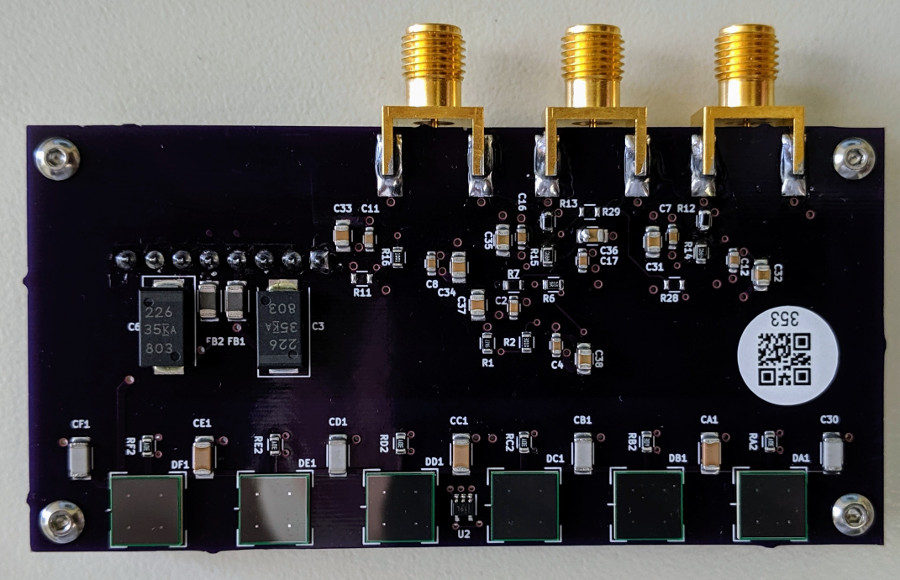}
\caption{Prototype preamplifier board. For this prototype three outputs were tested whereas the flight version will only have two.}
\label{fig:SiPM}
\end{subfigure}
\caption{\ac{tof} counter and readout electronics}
\label{fig:TofPaddle}
\end{figure}

\subsection{Thermal System}
Crucial to the operation of the tracker system (\ref{sec:track}) is the low-temperature operation of the Si(Li) detectors ($< \SI{-40}{\celsius}$).  To achieve this a newly-developed \ac{ohp} system will be used, which will transport the heat generated in the Si(Li) detector electronics and infrared inputs from the surrounding \ac{tof} system to a radiator that will dissipate the heat to space.

This system will be composed of 36 capillary loops servicing all 10 tracker layers and connected in series.  Inside the loops a two-phase working fluid will absorb the heat and, through passive circulation and self-oscillation excited in the interconnected loops, carry it to the radiator.  These passive driving forces allow for a low-power, and large-scale thermal system. To improve the robustness of the system, some active components have been added (heaters, valves and a mechanical pump).

As an innovative system in long-duration ballooning, the thermal system has been evaluated on scaled-down and full-scale models which have been tested in thermal chambers and on stratospheric balloon flights. A more complete description of the development and testing of the \ac{ohp} system is given in \cite{fuke2016}.

\subsection{Simulation, Reconstruction and Data Analysis}
\begin{figure}
\centering
\includegraphics[width=0.8\textwidth]{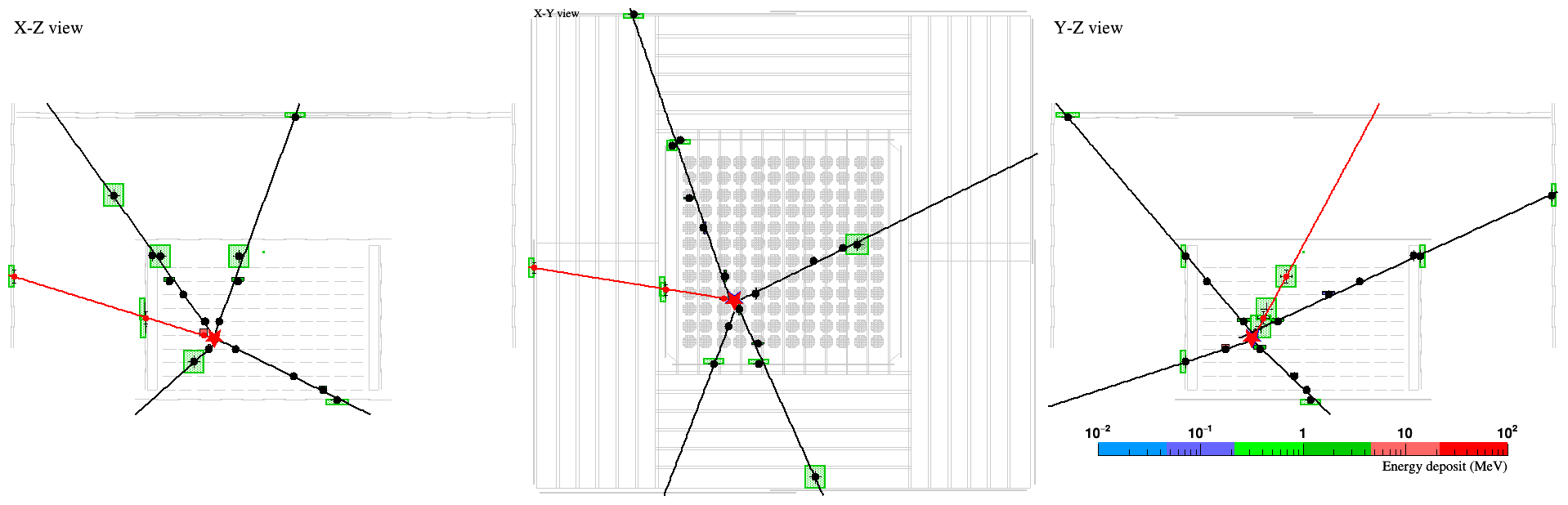}
\caption{An antideuteron event, showing reconstructed tracks.  The red line and dots are the reconstructed primary, the black line and dots are secondary particles.  The color scale shows the amount of energy deposited and the area the error on the position of the energy deposition.}
\label{fig:RecEvent}
\end{figure}

The performance of the \ac{GAPS} detector is being determined using dedicated GEANT4 simulations \cite{geant4}.  These are also being used to develop event reconstruction and classification algorithms.  A sample event is shown in \cref{fig:RecEvent} which shows the effective track identification and reconstruction (the current status of this work is presented in \cite{munni2019}).  Using the latest model of the detector and Monte Carlo tracking (i.e. not using reconstructed events), the acceptance for antideuterons has been calculated, peaking at over \SI{1}{m^2 sr}.

\section{Conclusions}
\ac{GAPS}, a balloon-borne experiment to search for low-energy antinuclei in cosmic rays as a signal of dark matter annihilation or decay, is on course for its first flight in the austral summer of 2021-22.  Using the novel exotic atom technique, it will provide the most sensitive low-energy ($<\SI{0.25}{GeV/nucleon}$) search for antinuclei to date, using a method that is complementary to the magnetic spectrometer measurements by BESS and AMS-02.  Significant progress has been made on all aspects of the design, including the fabrication of low-noise, high-temperature, large-area Si(Li) detectors, the production of a large-area, high-precision plastic scintillator time-of-flight and the required mechanical, electrical and computational support.

\acknowledgments{This work is supported in the U.S. by NASA APRA grants (NNX17AB44G, NNX17AB45G, NNX17AB46G, and NNX17AB47G), in Japan by JAXA/ISAS Small Science Program FY2017, and in Italy by Istituto Nazionale di Fisica Nucleare (INFN) and by the Italian Space Agency through the ASI INFN agreement n. 2018-28-HH.0: "Partecipazione italiana al GAPS - General AntiParticle Spectrometer". R.A. Ong receives support from the UCLA Division of Physical Sciences. K. Perez receives support from the Heising-Simons Foundation and Alfred P. Sloan Foundation. F. Rogers is supported through the National Science Foundation Graduate Research Fellowship under grant 1122374. P. von Doetinchem receives support from the National Science Foundation under award PHY-1551980. H. Fuke receives support from JSPS KAKENHI grants JP26707015, JP17H01136, and JP19H05198. M. Kozai receives support from JSPS KAKENHI grant JP17K14313. S. Okazaki receives support from JSPS KAKENHI grant JP18K13928. Y. Shimizu receives support from Sumitomo Foundation grant. }

\begin{acronym}
\acro{ohp}[OHP]{oscillating heat pipe}
\acro{tof}[TOF]{time-of-flight}
\acro{GAPS}[GAPS]{General Antiparticle Spectrometer}
\acro{ISM}[ISM]{interstellar medium}
\acro{sipm}[SIPM]{silicon photomultiplier}
\acro{fwhm}[FWHM]{full width half maximum}
\acro{asic}[ASIC]{application specific integrated circuit}
\acro{slider}[SLIDER]{Silicon LIthium DEtectors Readout}
\end{acronym}

\end{document}